\documentstyle[12pt]{article}
\title{ Cosmology, Particle Physics, and Superfluid 3He }
\author{G.E. Volovik\\
Low Temperature Laboratory, Helsinki University of
Technology \\ FIN-02150 Espoo, Finland \\ and \\  L.D. Landau Institute for
Theoretical Physics \\
   Kosygin Str. 2, 117940 Moscow, Russia.}
\begin{document}
\maketitle
\begin{abstract}
{
Many direct parallels connect superfluid $^3$He with
the field theories describing the physical vacuum, gauge fields
and  elementary fermions. Superfluid $^3$He exhibits a
variety of topological defects which can be detected with
single-defect sensitivity. Modern scenarios of defect-mediated
baryogenesis can be simulated by the interaction of the
$^3$He vortices and domain walls with fermionic quasiparticles.
Formation of defects in a symmetry-breaking phase transition in the early
Universe, which could be responsible for large-scale structure formation and for
microwave-background anisotropy, also may be modelled in the laboratory. This is
supported by the recent observation of vortex formation in
neutron-irradiated
$^3$He-B where the "primordial fireball" is formed in an exothermic nuclear
reaction. }
\end{abstract}

\section{INTRODUCTION}

There are many similarities between superfluid $^3$He and particle
physics stemming from the fact that both systems are described by the quantum
field theory. The superfluid $^3$He is a unique system among the other condensed
matter, because it has the maximum symmetry breaking, which  can be compared
only  with the vacuum in the elementary particle physics, and due to the rich
set of the fermionic and bosonic excitations interacting with gauge-like fields
of numerous collective modes \cite{Exotic,VolovikVachaspati}. Most of
the common phenomena are yet to be confirmed in the particle physics
while they are quite routinely observed or can be investigated in $^3$He. These
include the variety  of topological defects with their potential astrophysical
and cosmological consequences.

Till now in particle physics only the perturbative features of the modern
quantum field theories have been put to the test and it would be worthwhile to
have a better understanding of the non-perturbative aspects. Such
an understanding may be crucial, for example, for testing the hypothesis that
the baryon number of the Universe was produced (baryogenesis) during the
electroweak or other phase transition in the early Universe. To gain some
intuition, we need such condensed matter system as  superfluid $^3$He, which has
the closest similarity with the physical vacuum. This makes the superfluid
$^3$He a working laboratory for modelling different processes which can occur in
the physical vacuum and in Universe.

\section{VORTICES and COSMIC STRINGS}
\subsection{Topological defects}
The complicated topology of the $^3$He vacuum results in a variety
of defects in $^3$He, such as monopoles, hedgehogs,
boojums, solitons, domain walls, textures, vortices, which have
many direct analogies with topological defects in quantum vacuum. In
particular, due to the similar symmetry breaking in $^3$He-A and in the
electroweak vacuum\cite{VolovikVachaspati}, some of four  vortices observed
in $^3$He-A\cite{AphDiag} share the common properties with the $Z-$ and $W-$
strings of the electroweak model\cite{nambu,tv}. There are also
combinations of topological defects of different dimensionality found in
Helsinki
rotating cryostat: In $^3$He-A it is the vortex sheet -- the soliton
plane filled with the vortex lines \cite{VortexSheet}, which topology is
similar to that of Bloch lines within the Bloch wall in ferromagnets. In
$^3$He-B it is the soliton  terminating on the line
defect\cite{WallTerminatingOnString}, this combined object  is topologically
similar to the cosmic wall terminating on cosmic string\cite{Review2}.

\subsection{Phase transitions in vortices and strings.} Several phase
transitions related to the quantized vortices have been observed in Helsinki.
(i) Spontaneous breaking of continuous symmetry in the vortex core was
experimentally verified in $^3$He-B: a new Goldstone mode, related to the twist
of the deformed vortex core, has been excited in NMR
experiments\cite{CoreAsymmetry}.  In cosmic strings the analogous breaking of
$U(1)$ symmetry in the core  results in the superconducting current along the
twisted core\cite{Witten}. The instability towards the symmetry breaking in the
core can be triggered by fermions living in the core, see \cite{Naculich1995}
for cosmic strings and \cite{MakhlinVolovik1995} for vortices.

(ii) The textural phase transitions with the change of the topological
invariants
have  been identified in rotating $^3$He-A\cite{AphDiag}. In some of them the
transformation of the vortex  can occur by motion of point defects along
the vortex line\cite{SalomaaVolovik1987}. The relevant point defect can be
the monopole (analog of Dirac magnetic monopole) or the
hedgehog (analog of t'Hooft-Polyakov monopole). In the future this
scenario of the phase transiton will be tested in detail.

\subsection{Half-quantum vortices} The
vortices with the fractional winding number $N=1/2$ \cite {VolovikMineev1976}
are the counterpart of  Alice strings, which appear in some models of particle
physics (see eg \cite{McGraw}). The particle travelling around some type of the
Alice string changes its electric charge to the opposite
\cite{Schwarz1982}.  In  $^3$He-A, the analogous effect is the reversal of the
spin of the quasiparticle upon circling the 1/2 vortex. This behavior results
also in the peculiar Aharonov-Bohm effect, which has been discussed for  1/2
vortices in $^3$He-A \cite{Khazan1985,SalomaaVolovik1987} and  has been modified
for the cosmic Alice strings in \cite{March-Russel1992,Davis1994}.
The 1/2-vortex can also occur in the $d$-wave
superconductors\cite{Geshkenbein1987} and recently it has been identified in
high-temperature superconductor \cite{Kirtley1996}.

\section {PROBING OF VACUUM}
\subsection {Vacuum instability}  The instability of
the physical vacuum can occur in strong electric or gravity fields. The
former can be realized in collision of two heavy nuclei\cite{Gershtein} (see
also Refs.\cite{Grib,Calogeracos}), while the latter can happen in the vicinity
of the black hole. The analog of the former instability is realized in the
moving $^3$He-B, if its velocity reaches the Landau value: The
creation of quasiparticles in supercritical regime, detected in
vibrating wire experiments\cite{Pickett}, is similar to the creation of the
electron-positron pairs by strong field.

\subsection{Gravity and 3He} In many cases the
motion of the elementary excitations in condensed matter corresponds to the
motion  of the particle in the effective gravity field represented by the metric
tensor \cite{UnruhSonic,EffectiveGravity,Exotic,Ilinski}. In superfluid 3He the
gravity field is simulated by the order parameter texture. This
condensed matter analog of Sakharov's effective gravity\cite{Sakharov} allows to
model several phenomena related to strong or quantum gravity: (i) In the
effective gravity the cosmological constant is many orders of magnitude larger
than the experimental upper limit\cite{Weinberg}. Superfluid
$^3$He possibly gives some hint how to resolve this
paradox\cite{EffectiveGravity,Exotic}. (ii) The problem of the black hole
entropy and quantum radiation\cite{Hawking} can be modelled by moving texture
(soliton or interface between $^3$He-A and $^3$He-B). When the velocity of the
texture exceeds the critical value (which is rather low and is easily achieved
in the 3He experiment) the metric resembles that of the black hole: the event
horizon appears\cite{JacobsonVolovik}. This is the  realization of the idea of
Unruh who introduced the event horizon in condensed
matter\cite{UnruhSonic}. This should help to understand such problems in
quantum gravity as possibility of the nonunitarity of quantum state and  the
behavior of the vacuum in strong gravity. The black hole analogy was also
discussed  for closed string
\cite{Copeland} and vortex\cite{VortexBlackHole}.

\subsection{Homogeneously precessing vacuum} Nonlinear spin-coherent dynamics
of the $^3$He-B vacuum discovered in \cite{Borovik-Fomin} is the subject of
intensive investigations. Several dynamical vacua, precessing in the
applied magnetic field, have been found and identified in NMR experiments (see
the latest Refs.\cite{1/2spin,Bunkov}). The analogy with the nonperturbative
dynamics of the physical vacuum is to be found.

\subsection{Other vacuum effects} Vacuum polarization and zero charge
phenomena have their counterpart in $^3$He \cite{Exotic}. The vacuum pressure
(Casimir effect) and vacuum friction (quantum radiation from the mirror moving
in the physical vacuum, see Refs. in \cite{Barton}) are reproduced by the moving
interface between $^3$He-A and $^3$He-B \cite{Casimir}.

\section{VORTEX MOTION vs BARYOGENESIS}

The popular hypothesis of the baryogenesis is that the baryon number of the
Universe was produced during the electroweak phase
transition in the early hot Universe.  According to modern scenarios
\cite{Review2,Turok} the baryogenesis was
mediated by  topological defects. Similar situation
occurs in superfluid $^3$He and superconductors, where the dynamics of
vortices and domain walls leads to the production of the fermionic charges. In
this Section we discuss   the "momentogenesis" -- production of the
fermionic linear momentum during the vortex motion. This leads to the
additional nondissipative force on the vortex which competes with conventional
Magnus force.

\subsection{Three forces on moving vortex}
The dynamics of the vortex is governed by interaction of 3 object: (1)
Superfluid {\it vacuum} moving with the velocity ${\bf v}_s$. (2) The system
of excitations ({\it matter}) which form the normal component of the liquid. It
moves with the velocity ${\bf v}_n$ (heat bath or normal velocity). (3) The
vortex moving with the velocity ${\bf v}_L$. The vortex also carries the
excitations localized in the vortex core \cite{Caroli}.

As a result the equation for the balance of forces acting on the moving
vortex contains 3 nondissipative forces and the friction
one\cite{KopninVolovik1995}:
$${\bf F}_{\rm M}+{\bf F}_{\rm I}+
{\bf F}_{\rm sf} + D({\bf v}_n-{\bf v}_L)=0
\eqno(4.1)
$$
Here $D$ is the parameter of the friction force, which arises when the vortex
moves with respect to the heat-bath. The nondissipative forces contain (i)
the conventional Magnus force on the vortex moving with respect to the
superfluid vacuum
$$
{\bf F}_{\rm M}=  \rho N \vec \kappa\times({\bf v}_L-{\bf v}_s)~~.
\eqno(4.2)
$$
Here $N$ is the vortex winding number,
$\kappa$ is the circulation quantum, $\rho$ the density of the liquid.

(ii) Iordanskii force is the result of the Aharonov-Bohm effect,
experienced by excitations in bulk liquid, when they interact with the
vortex\cite{Sonin}:
$$
{\bf F}_{\rm I }=N\vec \kappa\times {\breve \rho}_n(T)
({\bf v}_s-{\bf v}_n) ~~ .\eqno(4.3)
$$
Here ${\breve \rho}_n(T)$ is the (tensor) density of the normal
component. This effect is analogous to the gravitational
Aharonov-Bohm topological effect discussed for the spinning
cosmic string, ie for the strings with an angular
momentum\cite{CausalityViolation,3forces}.

(iii) The so called spectral-flow force is responsible for
the "momentogenesis" which we discuss below
$$
{\bf F}_{\rm sf }=NC(T)\vec \kappa\times  ({\bf v}_n-{\bf v}_L) ~~.
\eqno(4.4)
$$
The parameter $C(T)$ is determined by the anomalous dynamics of
fermions in the core\cite{KopninVolovik1995,Stone}.

\subsection{Baryogenesis and axial anomaly.}

There is a close connection between the spectral flow force and the
nonconservation of the baryon number $N_B$  in the presence of the  $SU(2)$ and
$U(1)$ field strengths ${\bf F}_{\mu \nu}$ and $F_{\mu \nu}$
$$
\partial_t N_B
= {{N_F} \over {32\ \pi^2}} \int d^3r\left (
- {\bf F}_{\mu \nu}  {\tilde {\bf F}}^{\mu \nu} +   F_{\mu \nu}
{\tilde F}^{\mu \nu} \right )
\eqno(4.5)
$$
Here $N_F$ is the number of families. The nonconservation of $N_B$ is the result
of the phenomenon of the axial anomaly predicted by Adler \cite{Adler1969} and
Bell and Jackiw\cite{BellJackiw1969}. The fields
${\bf F}_{\mu\nu}$ and $F_{\mu \nu}$ can be generated in the core of the
cosmic strings evolving in the expanding
Universe \cite{Review2,Turok,tvgf,jgtv}.

\subsection{Momentogenesis in 3He-A.}

Similar nonconservation of the fermionic charge occurs in
$^3$He-A. The $^3$He-A quasiparticles in the vicinity of the point gap nodes are
chiral: like neutrino they are either left-handed  or right-handed
\cite{Exotic}.  For such gapless chiral fermions the  Adler-Bell-Jackiw
anomaly of the kind in Eq.(4.5) takes place. The "chiral charge"
of quasiparticles $N_{ch}$ is not conserved in the presence
of the {\it electric}  and {\it magnetic}  fields
${\bf E}=\partial_t {\bf A}$, ${\bf
B}=\vec\nabla\times {\bf A}$:
$$
\partial_t N_{ch}
=     {1\over {2\pi^2}}\int d^3r ~\partial_t {\bf A} \cdot
                  (\vec  \nabla \times {\bf A}  \, \, )  ~~.
\eqno(4.6)
$$
In  $^3$He-A  the vector potential ${\bf A}$ acting
on the quasiparticles is simulated by the  ${\hat{\bf l}}$ texture,  ${\bf
A}=p_F{\hat {\bf l}}$, where ${\hat {\bf l}}$ is a unit vector in the direction
of the gap node and $p_F$ the Fermi momentum.

For us it is important that the left quasiparticle carries the linear momentum
$p_F{\hat {\bf l}}$, and the equal momentum is carried by the left
quasihole. As a result the counterpart of the axial anomaly in condensed matter
leads to the net product of the fermionic linear momentum ${\bf P}$ in the
time-dependent texture:
$$
\partial_t  {\bf P}=
     {1\over {2\pi^2}}\int d^3r~ p_F\hat {\bf l} ~(\partial_t {\bf A}
            \cdot (\vec  \nabla \times {\bf A}  \, \, ))  ~~.
\eqno(4.7)
$$
Since the total linear momentum is conserved in condensed matter, the Eq.(4.7)
means that the momentum is transferred from the superfluid {\it vacuum}
to the {\it matter} (the normal component) in the presence of the
time-dependent texture.

\subsection{Momentogenesis by continuous vortex.}

The continuous vortex  moving in 3He-A generates both the ``electric''  and
``magnetic''  fields and thus represents the right texture,
which leads to the  momentogenesis. Integration of the anomalous momentum
transfer in Eq.(4.7) over the cross-section of the soft core of the moving
vortex gives the lost of the linear momentum\cite{Volovik1992}. This
corresponds to the force acting on the vortex in the form of  Eq.(4.4):
$$
{\bf F}_{\rm sf }=NC_0\vec \kappa\times  ({\bf v}_n-{\bf v}_L) ~~.
\eqno(4.8)
$$
The parameter $C(T)$ appears to be temperature independent, $C_0=m_3
p_F^3/3\pi^2$, and is close to the mass density $\rho$. The  continuous vortex
provides an example of the extreme spectral-flow force, which effectively
cancels the Magnus force in Eq.(4.2).

\subsection{Spectral-flow force on singular vortex.}

 For singular vortices in $^3$He-B and superconductors the spectral flow is
suppressed due to discrete character of the fermionic spectrum  in the vortex
core\cite{Caroli}. The anomalous exchange between the core fermions and the heat
bath now depends on the kinetics, determined  by the level spacing $\omega_0$
and the life-time $\tau$ of the core
fermions\cite{KopninCoAuthors,KopninVolovik1995,vanOtterlo1995,Stone}:
$$
C(T) \approx  C_0\Bigl[
1-{\omega_0^2\tau^2  \over 1+
\omega_0^2\tau^2}~\tanh{\Delta(T)\over 2T}~\Bigr] ~~.
\eqno(4.9)
$$
The extreme limit, $C(T)=C_0$, takes place when the interlevel distance can be
neglected compared to the life time: $\omega_0 \tau
\ll 1$, ie when the fermionic spectrum can be considered as continuous and the
spectral flow is not suppressed \cite{Volovik1993}.  In $^3$He-A and
in dirty enough superconductors this occurs at practically arbitrary
temperature, while for clean superconductors and for $^3$He-B this regime
appears only at high temperature (for $^3$He-B at $T>0.5~T_c$, see below).

The dynamics of the  core fermions also determines the friction parameter $D$
\cite{KopninCoAuthors,Stone}:
$$
D\approx \kappa C_0~\tanh{\Delta(T)\over 2T} {\omega_0\tau  \over
1+\omega_0^2\tau^2}~~.
\eqno(4.10)$$

\subsection{Experiment on vortex dynamics}
The forces on the vortex have been recently
measured in superfluid
$^3$He-B in  a broad temperature range \cite{Bevan1995}. The results
are expressed in terms of
$$
d_\parallel={D(T)\over \kappa\rho_s(T)}~,~d_\perp=
{C(T)-\rho_n(T)\over \rho_s(T)}\eqno(4.11)
$$
The experimental bell shape of  $d_\parallel (T)$  follows the temperature
dependence of $\omega_0\tau/ (1+\omega_0^2\tau^2)$ in the Eq.(4.10) with
$\omega_0\tau\gg 1$  at $T\ll T_c$ and $\omega_0\tau\ll 1$ close to $T_c$. The
temperature dependence of the observed $d_\perp(T)$ reproduces the Eq.(4.9): At
low $T$ the observed negative sign of $d_\perp(T)$   reflects the suppression of
the spectral flow according to Eq.(4.9): $C(T)\ll \rho_n(T)\ll \rho$.  The
observed upturn of $d_\perp(T)$   demonstrates an increase of $C(T)$
with increasing
$T$. The positive value of $d_\perp(T)$ at $T>0.5~T_c$ shows that at these $T$
the spectral flow already dominates:  $C(T)> \rho_n(T)$. And finally at higher
$T$, where
$\omega_0\tau\ll 1$, the experimental  $d_\perp(T)$ approaches 1, which shows
that $ C(T)\rightarrow \rho$, ie the spectral flow reaches its extreme limit.

\section{BIG BANG SIMULATIONS}

\subsection{Introduction} The string scenario of the
baryogenesis implies that the strings appear at some stage
of the expanding Universe. These strings could be also responsible for
large-scale structure formation and for microwave-background anisotropy of the
Universe\cite{Review2}. In a popular theory the topological defects can be
formed, if during the cooling of the Universe a symmetry-breaking phase
transition occurs at some critical temperature $T_c$ \cite{Zeldovich,Kibble}.
According to Zurek modification\cite{Zurek} of the Kibble
mechanism\cite{Kibble}, the topological defects appear
due to the nonequilibrium dynamics of the phase ordering. The nonequilibrium
processes take place because the cool-down through $T_c$ occurs with finite
rate, as a result the order parameter fluctuations above $T_c$ are quenched
below $T_c$.

\subsection{Advantages of 3He for simulations}

The physics of the nonequilibrium phase transition in expanding Universe is the
same as in condensed matter, and thus may be modelled in the laboratory
where the cooling rate can be controlled \cite{Zurek}.  Laboratory experiments
intended to test different theories of cosmological defect formation have
recently been conducted in liquid crystals \cite{nematic} and superfluid $^4$He
\cite{helium}. The superfluid phases of liquid $^3$He open a new page in these
Big-Bang simulations. $^3$He has several advantages over other systems:
(i) It is more closely resembles the complicated physical vacuum. (ii) It has a
variety  of bosonic and fermionic elementary particles which are very important
for the physics of the evolution of the topological defects.  The fermionic
quasiparticles can be detected with sensitive vibrating wire
detector\cite{DarkMatter}. (iii) The superfluid $^3$He exhibits several phase
transitions\cite{VolWol}, this allows us to investigate the defects formation
after the 1-st and the 2-nd order transitions. (iv) We can investigate formation
of different types of the topological defects, some of them can be detected with
NMR methods with single-defect sensitivity
\cite{Nucleation}.

(v) Of particular relevance to the
quench experiment is the fact that liquid $^3$He can be locally efficiently
heated with thermal
neutrons\cite{Schiffer-Osheroff,DarkMatter,BigBangNature1,BigBangNature2}.
Thermal neutrons produce the `primordial fireball' due to the nuclear reaction n
+ $^3_2$He = p +
$^3_1$H + 0.76MeV.  In Helsinki experiments \cite{BigBangNature1} it was found
that the subsequent rapid cooling of the locally overheated liquid through the
superfluid transition $T_c$  results in the formation of vortex rings.  If the
$^3$He-B container rotates, sufficiently large vortex rings grow under the
influence of the Magnus force, escape from the heated bubble, and are detected
individually with nuclear magnetic resonance.

\subsection{Primordial fireball by nuclear reaction}

In Helsinki experiment \cite{BigBangNature1} the neutrons, produced with a
paraffin moderated Am-Be source of 10 mCi activity, are incident upon
the $^3$He sample container. At the minimum distance of 22 cm between source and
sample, $\nu \approx 20$ neutrons/min are absorbed by the $^3$He liquid. For
thermal neutrons the  mean free path is 0.1 mm in liquid $^3$He and thus all
absorption reactions occur close to the walls of the container. Since the
incident thermal neutron has low momentum, the 573 keV proton and 191 keV triton
fly apart in opposite directions, producing 70 and 10 $\mu$m long ionization
tracks, respectively. The subsequent charge recombination yields a `primordial
fireball' -- a heated region of the normal liquid phase. This process was
discussed in details in connection to the $^3$He-A $\rightarrow$ $^3$He-B
transition mediated by neutrons \cite{Schiffer-Osheroff}: the fireball may lead
to the formation of the critical bubble of the $^3$He-B phase in the supercooled
$^3$He-A. As in \cite{Schiffer-Osheroff} we assume for simplicity that the
fireball has a spherically symmetric shape.

This bubble of normal fluid cools by the diffusion of quasi-particle
excitations out into  the surrounding superfluid with a diffusion constant $D
\approx v_F l$ where $v_F$ is their Fermi velocity and $l$ the mean free path.
The difference from the surrounding bulk temperature $T_0$
as a  function of the  radial distance $r$ from the centre of the bubble is
given by $T(r,t) - T_{0} \approx (E_0/ (4 \pi D t
)^{3/2}C_v) \exp
( -r^2/ 4Dt   )$
where $E_0$ is the energy deposited by the neutron as heat and $C_v$ the
specific heat. The maximum
value  of bubble radius $R_{b}$ with fluid in the normal phase,
$T(r)>T_c$, is
$$
 R_{b} \sim  (E_0/ C_v T_c)^{1/3} (1-T_{0}/T_c)^{-1/3}~,\eqno(5.1)
$$
which is typically of order 10 $\mu$m.
The bubble cools and shrinks away rapidly with the characteristic  time
$\tau_Q \sim R_b^2/D
\sim 10^{-6}$ s.

\subsection{Defect formation by quench}

The distance $\xi_v(t)$ between the defects in the process of the phase ordering
after the quench is of primary interest. The real defects appear at the moment
$t^\star$ when they can be distinguished from the background fluctuations of the
order parameter. The density of the defects at $t^\star$ (the so called initial
vortex density) was recently the subject of controversy. In the original Kibble
scenario $t^\star$ was identified with the moment when the system cools below
the Ginzburg temperature $T_G$ above which the thermal fluctuations prevail, ie
$T(t^\star)=T_G$. This implies that the initial distance between the vortices is
the coherence length at $T_G$, i.e.
$\xi_v(t^\star)=\xi(T_g)=\xi_0 /\sqrt {1-(T_G/T_c)}$, and does not depend on
the cooling time $\tau_Q$. In superfluid $^3$He the critical fluctuation region
is extremely narrow, $1-(T_G/T_c)\sim (a/\xi_0)^{4}$, because the  zero
temperature coherence length $\xi_0\sim 0.1~\mu$m is two orders of magnitude
larger than the interatomic spacing $a$\cite{VolWol}. As a result 
$\xi(T_g) \sim\xi_0(\xi_0/a)^{2}$ exceeds the bubble size $R_b$,
which means that no vortices can be created in this scenario.

\vspace*{7cm}
{\small Fig.\,1. \,  Two length scales characterizing the nonequilibrium
phase transtion into the ordered phase: coherence length $\xi(t)$ and
intervortex
distance in the infinite cluster $\xi_v(t)$. At $t>t^\star$ one has
$\xi_v(t)>\xi(t)$ and the vortices become well defined in the ordered phase.
}
\vspace{2mm}

In the alternative theory put forward by Zurek \cite{Zurek,ZurekReview} the
initial vortex density is essentally determined by the cooling rate
$1/\tau_Q=\partial_t T/T_c$ (if $t=0$ is the moment of phase transtion one has
$1 -T(t)/T_c=t/\tau_Q$). To get the idea of this theory let us consider an
oversimplified model, in which the normal state above $T_c$ is represented as an
infinite cluster of vortices  with the intervortex distance $\xi_v$
determined by the coherence length $\xi(T)$. The equilibrium state below $T_c$
does not contain an infinite cluster, but in the nonequilibrium phase transition
this cluster persists even in the ordered phase.  When
$T\rightarrow T_c$  the intervortex distance $\xi_v(t)$ diverges together with
the thermodynamic coherent length, but at some moment $t_0$ (see Fig.1) the
velocity $\partial_t\xi_v$ of defects approaches the limiting speed of light
$c$. At $t>t_0$ the distance between the vortices $\xi_v(t)$ can increase only
with the speed $c$. As a results in nonequilibrium the system has
two length scales: $\xi_v(t)$ characterizes the infinite cluster, while the
coherence length $\xi(t)$ is determined by the other degrees of freedom
(including the small closed loops).

After transition into the ordered state  (which occurs at $t=0$) the
$\xi(t)$  decreases and $\xi_v(t)>\xi(t)$ at $t>t^\star$ (see Fig.1).
Starting from that moment $t^\star$ the vortices in the infinite cluster become
well defined. The initial distance between the vortices in this model,
determined from the equation  $\xi_v(t^\star)=\xi(t^\star)$, is
$\xi_v(t^\star)\sim\xi_0(\tau_Q/\tau_0)^{1/3}$, where
$\tau_0=\xi_0/c$.  In $^3$He  the limiting speed $c$ relevant for the order
parameter defects is the velocity of propagation of the order parameter, which
is of order of spin-wave velocity: $c\sim v_F\sqrt {1-(T/T_c)}$. Then one has
$$\xi_v(t^\star)\sim\xi_0(\tau_Q/\tau_0)^{1/4}~,\eqno(5.2)$$
where $\tau_0=\xi_0/v_F$. With $\tau_Q$ from Sec.(5.3) this gives the initial
distance between the defects $\xi_v(t^\star)\sim 1\mu$m. This is well
within bubble radius $R_b$, ie after cooling through $T_c$ the bubble should
contain many well defined vortex lines.

\subsection{Evolution of the string network}

Below $T_c$ the  vortex spaghetti is nonequilibrium and thus finally decays. At
high temperature used in \cite{BigBangNature1} the decay occurs
within the former fireball. At low $T$ of \cite{BigBangNature2}
the friction is small and  vortices have time to propagate into the bulk
liquid before they collapse. The formed vortices can be effectively
stabilized in the rotating container which produces the superfluid velocity
$v_s =\Omega R$ near the wall (here $R$ is the container  radius
and $\Omega$ the rotation velocity). If  the radius of the
vortex loop exceeds the value $r_\circ(v_s) = ( \kappa/ 4\pi v_s)
\ln{r_\circ/\xi}$, where $\kappa=\pi\hbar/m_3$ is the circulation quantum, the
loop expands under the action of the Magnus force. An expanding vortex ring
eventually results in a rectilinear  vortex line which is pulled to the centre
of the container, where it is detected by NMR. This allows us to measure the
number of loops in the cluster with $\xi_v > r_\circ(v_s)$.

For the velocities $v_s$ used in the experiments, the radius
$\xi_v(t^\star)$ of the initial loops is smaller than the critical radius
$r_\circ(v_s)$, that is why the formed vortex network remains in the bubble and
the vortex cluster decays according to the general properties of the strings
dynamics. If in this process the $\xi_v(t)$  reaches  $r_\circ(v_s)$,
the vortices  escape and are detected. Let us discuss this in more detail.

During the decay of the cluster the network structure remains scale invariant,
while  $\xi_v(t)$ gradually increases \cite{Review2}.  The number of loops
$n(l)$ per unit length and unit volume with line lengths $l > \xi_v$ is given
\cite{VachaspatiVilenkin}  by $n(l)=C\xi_v^{-3/2}l^{-5/2}$ where $C \sim
0.3-0.4$.  The network is also characterized by the  average straight-line
dimension $\cal D$ of a loop: ${\cal D} =\beta (l\xi_v)^{1/2}$, which minimal
value is ${\cal D}_{\rm min}=\alpha \xi_v(t)$ (from numerical simulations one
finds that $\beta$ and $\alpha$ are close to  unity). In terms of ${\cal D}$ the
loop size distribution, $n({\cal D})\;d{\cal D} \approx 2C
~d{\cal D} / {\cal D}^4 $, does not depend on $\xi_v$:
The evolution of the network leads to an increasing lower cutoff of the
distribution  ${\cal D}_{\rm min}=\xi_v(t)$ while the  upper cutoff is the
diameter of the bubble ${\cal D}_{\rm max}=2R_b$. When the average radius of
curvature, determined by $\xi_v$,  exceeds $r_\circ (v_s)$ the vortices start to
expand. Thus the number of detected vortices per one Big-Banh event,
${\cal N}(v_s)$,  is  the  number of loops with
$ r_\circ(v_s) < {\cal D} < 2R_b$ in the bubble volume $V_b$:
$$
{\cal N}(v_s)=V_b\int_{r_\circ }^{2R_b}
d{\cal D}~
n({\cal D})= {\pi C\over 9} \left[\left({{2R_b} \over {r_\circ(v_s)}}\right)^3
-1\right]   \eqno(5.3)$$

The Eq.(5.3) shows that there should be the critical velocity $ v_{cn}$, below
which no vortices can be detected: the vortices can
be extracted only if $r_\circ(v_s)<2R_b$, which gives
$$v_{cn}  = (\kappa  / 8\pi  R_b) \; \ln  (R_b/\xi)~. \eqno(5.4)$$
According to Eq.(5.1) $v_{cn}$ has the temperature dependence
$v_{cn} \propto (1-T_{0}/T_c)^{1/3}$.
In terms of $ v_{cn}$ one obtains the universal curve
$$
{\cal N}(v_s/v_{cn})  = {\pi C\over
9}\left[\left({v_s\over v_{cn}}\right)^3 -1\right] \eqno(5.5)$$

\subsection{NMR measurement of vortex formation}

In the inset of Fig.~2 two NMR absorption records are shown, starting from the
moment when the neutron source is placed in position. The absorption events,
which lead to formation of rectilinear  vortices, are visible as distinct steps.
The step height gives the number of vortex lines per event. The number of
vortex lines created per unit time $\dot N=\nu \cal N$ (main frame of Fig.~2)
reproduces the theoretical $v_s^3$ dependence in Eq.(5.5) and is of the
correct order of magnitude. The Fig.~3 demonstrates the universality feature of
$\dot N$: The measuring conditions depend on temperature, pressure, and magnetic
field, but  all this dependence is contained in
the single parameter, the critical velocity $v_{cn}$. The temperature dependence
and the order of magnitude of $v_{cn}$ are in agreement with Eq.(5.4). This
$v_{cn}$ is much smaller  than the critical velocity $v_c$ at which a vortex is
created at the container wall in the absence of neutrons\cite{Nucleation}.

\vspace{2mm}

\vspace*{6.5cm}
{\small Fig.\,2. \, Formation of vortices during neutron irradiation.
{\it Inset:} Change in NMR absorption as a function of running time $t$ at
high and low
rotation velocity. Each step corresponds to one neutron absorption event. The
height of the step
gives the number
of vortex lines created and is denoted by the adjacent
number. {\it Main
frame:} The rate
$\dot N$ at which  vortex lines are created during neutron irradiation,
plotted as a function of the normalized superflow velocity $v_s/v_{cn}$.
Below the critical threshold at $v_{cn}$ the rate vanishes while above
$v_{cn}$ the rate follows
the fitted cubic dependence $\dot N =\nu {\cal N}(v_s/v_{cn})$,
where $\nu \approx 20$ neutrons/min is the neutron flux and $\cal N$ is
given by Eq.(5.5) with the fitting parameter $C\sim 0.2$.
}

\vspace*{6.5cm}

{\small Fig.\,3. \, The rate
$\dot N$ as a
function  of $v_s^3$ demonstrates the scale invariance of
the vortex formation in different
conditions.
In agreement with Eq.~(5.5) the dependence on temperature,
pressure, and magnetic field is contained in the critical velocity
$v_{cn}$. The latter is obtained as the intercept of the fitted lines with
the horizontal axis. The common intercept with the vertical axis gives $\nu \pi
C/9$.
}
\vspace{2mm}

\subsection{Formation of other defects}

There is an experimental evidence of the other type of the topological defects,
formed in the mini-Big-Bang events. This is the spin-mass vortex (with the mass
current and spin current circulating together around the vortex line), which is
also the termination line of the soliton\cite{WallTerminatingOnString}. Due to
the mass flow circulation the spin-mass vortex is influenced by the Magnus force
in the same manner as conventional (mass) vortex, and can be also extracted
under rotation. The events interepreted as
formation of  spin-mass vortices occur rather rare, as one can expect for
the combined defects.

Another object which should form in the rapid quench is the interface between
$^3$He-B and $^3$He-A, however without the bias field the A-B interfaces decay
within the Big-Bang volume. Situation changes in the supercooled $^3$He-A: the
energy difference between  $^3$He-A and $^3$He-B gives the required bias
for the growth of the interfaces, which finally results in A$\rightarrow$B
transition. This provides an  explanation of the
observed\cite{Schiffer-Osheroff}  collapse of the supercooled
$^3$He-A by neutron and $\gamma$ irradiation, which is alternative
to the "baked Alaska" scenario \cite{Leggett1984}.

In the future the formation of half-quantum vortices (Alice strings) and point
defects (monopoles) in the Big-Bang events will be studied.

\subsection{Discussion}

The experiment demonstrates that vortices are created in a rapid quench to the
superfluid state. The measured number of vortices created as a function of
superfluid velocity $v_s$ is consistent with the Vachaspati-Vilenkin scaling law
\cite{VachaspatiVilenkin} discussed for cosmic strings. This is in favour of
the Zurek modification of the Kibble mechanism of defect formation. On the other
hand it is difficult to find some other mechanism of the defect nucleation by
neutron irradiation, which could be responsible for all the found phenomena,
including that  observed at low $T$ \cite{BigBangNature2}. It is believed
now that the  $v_s$ itself is not the reason of the vortex formation:  $v_s$
mainly allows the formed vortices to
escape, to expand and finally to reach a stable state in the rotating container
so that they can be detected one by one. With increasing  $v_s$,
smaller loops, which represent an earlier stage in the evolution of the network,
are extracted. One can expect that when the smallest possible size of loops is
reached, which is limited by the initial inter-defect distance $\xi_v(t^\star)$,
the cubic scaling law will be violated. At the moment, the smallest size of the
loops extracted at the highest rotation velocity is above though close to
$\xi_v(t^\star)$.

\section{CONCLUSION}

There is a close connection between superfluid $^3$He and high energy
physics. Many nonperturbative processes in particle physics and cosmology can be
modelled in $^3$He. On the other hand some effects observed in
$^3$He still wait for their analogs in physical vacuum. Also the $^3$He is a
very pure system and can serve for the particle physics
experiments, eg as the dark matter detector\cite{DarkMatter}.

\end{document}